\documentclass[12pt]{article}
\usepackage{graphicx}

\newcommand{\beq}{\begin{equation}}
\newcommand{\eeq}{\end{equation}}
\newcommand{\bea}{\begin{eqnarray}}
\newcommand{\eea}{\end{eqnarray}}
\newcommand{\half}{\frac{1}{2}}
\newcommand{\dc}{^\circ{\rm C}}
\newcommand{\dT}{\Delta T}
\newcommand{\ms}{~{\rm ms}^{-1}}

\begin{document}
\begin{center}
\noindent{\Large The Effects of Temperature, Humidity and Barometric Pressure
on Short Sprint Race Times}
\vskip 1cm
J.\ R.\ Mureika\\
{\it 
Department of Physics\\ Loyola Marymount University\\
1 LMU Drive, MS-8227\\ Los Angeles, CA~90045}
\vskip .5cm
Email: jmureika@lmu.edu\\
\end{center}
\vskip 2cm 

\noindent
{\footnotesize
{\bf Abstract} \\
A numerical model of 100~m and 200~m world class sprinting performances is modified
using standard hydrodynamic principles to include effects of air temperature,
pressure, and humidity levels on aerodynamic drag.  The magnitude of the
effects are found to be dependent on wind speed. This implies that differing 
atmospheric conditions can yield slightly different corrections for the same 
wind gauge reading.  In the absence of wind, temperature is found to induce 
the largest variation in times (0.01~s per $10\dc$ increment in the 100~m), 
while relative 
humidity contributes the least (under 0.01~s for all realistic conditions for
100~m).  
Barometric pressure variations at a particular venue can also introduce 
fluctuations in performance times on the order of a 0.01~s for this race.  
The combination 
of all three variables is essentially additive, and is more important for
head-wind conditions that for tail-winds.  As expected, calculated corrections 
in the 200~m are magnified due to the longer duration of the race.  
The overall effects of these 
factors on sprint times can be considered a ``second order'' adjustment to 
previous methods which rely strictly on a venue's physical elevation, but
can become important in extreme conditions.
}

 \pagebreak

\section{Introduction}
Adjusting athletic sprinting races for atmospheric drag effects has been
the focus of many past studies [1--18].
Factors influencing the outcome
of the races are both physiological and physical.  The latter has included
the effects of wind speed and variations in air density, primarily introduced
through altitude variations.  This is due in
part to the International Association of Athletics Federation's (IAAF) 
classification of ``altitude assistance'' for competitions at or above
1000~m elevation.  

For the 100~m, with the exception of a few references
it is generally accepted that a world
class 100~m sprint time will be 0.10~s faster when run with a 2$\ms$ 
tail-wind, where as a rise of 1000~m in altitude will decrease a race
time by roughly 0.03~s.  The figures reported in \cite{dapena} are
slight underestimates, although one of the authors has since re-assessed
the calculations and reported consistent values \cite{dapena2000} 
to those mentioned above.  A 0.135~s correction for this windspeed
is reported in \cite{behncke}, which overestimates the accepted value
by almost 40\%.

Wind effects in the 200~m are slightly more
difficult to determine, based on lack of data (the race is around a curve,
but the wind is measured in only one direction).
As a result, the corrections are less certain, but several
References \cite{jrm200,quinn1} do agree that a 2$\ms$ wind will provide
an advantage between 0.11-0.12~s at sea level.  Altitude adjustments
are much more spread out in the literature, ranging between 0.03-0.10~s
per 1000~m increase in elevation.  The corrections discussed in 
\cite{frohlich} and \cite{behncke} are mostly overestimates, citing
(respectively) 0.75~s and 0.22~s for a 2$\ms$ tail-wind, as well as 0.20~s and 
0.08~s for 1000~m elevation change (although this latter altitude correction
is fairly consistent with those reported here and in \cite{jrm200}.
Conversely, an altitude change of 1500~m is found to assist a sprinter
by 0.11~s in the study presented in \cite{quinn1}.

Several other factors not generally considered in these analyses, which are 
nonetheless crucial to drag effects, are air temperature, atmospheric 
pressure, and humidity level (or alternatively dew point).  There is a 
brief discussion of temperature effects in \cite{dapena}, but the authors 
conclude that they are largely inconsequential to the 100~m race.  However, 
the wind and altitude corrections considered therein have been shown to be 
underestimates of the now commonly accepted values, so a more detailed 
re-evaluation of temperature variations is in order.

This paper will investigate the individual and combined effects 
of these three factors, and hence will discuss the impact of {\it density
altitude} variations on world class men's 100~m and 200~m race times.  
The results suggest 
that it is not enough to rely solely on physical altitude measurements for 
the appropriate corrections, particularly in the 200~m event.  

\section{The quasi-physical model with hydrodynamic modifications}

The quasi-physical model introduced in \cite{jrm100} involves a set of
differential equations with five degrees of freedom,
of the form
\bea
\dot{d}(t) & = & v(t)  \nonumber \\
\dot{v}(t) & = & f_s + f_m - f_v - f_d~.
\label{quasi}
\eea
where the terms $\{f_i\}$ are functions of both $t$ and $v(t)$.  The first
two terms are propulsive (the drive $f_s$ and the maintenance $f_m$), while
the third and fourth terms are inhibitory (a speed term $f_v$ and drag
term $f_d$).  Explicitly,
\bea
f_s & = & f_0 \exp(-\sigma\; t^2)~, \nonumber \\
f_m & = & f_1 \exp(- c\;t)~, \nonumber \\
f_v &  =&  \alpha\; v(t)~, \nonumber \\
f_d &  =&  \half \left(1-1/4\exp\{-\sigma t^2\}\right) \rho A_d\;(v(t) - w)^2~.
\label{terms}
\eea
The drag term $f_d$ is that which is affected by atmospheric conditions.  The
component of the wind speed in the direction of motion $w$ is explicitly
represented, while additional variables implicitly modify the term by changing
the air density $\rho$.  

The motivation for using time as the control variable in the system of
equations (\ref{terms}) stems from the idea that efficient running of the
short sprints is an time-optimization problem.  This has been discussed
in the context of world records in Reference~\cite{keller}.

\subsection{The definitions of altitude}
In general, altitude is not a commonly-measured quantity when considering
practical aerodynamics, simply because the altitude relevant to such 
problems is {\it not} the physical elevation.  It is much easier to measure 
atmospheric pressure, which is implicitly determined by the altitude $z$ via
the differential relationship $dP(z) = -\rho\; g\; dz$, known as the
hydrostatic equation \cite{haughton}.  Thus, the pressure
$P(z)$ at some altitude $z$ may be obtained by integration assuming the
functional form of the density profile $\rho$ is known.  The air density 
can depend non-trivially on factors such as vapor pressure ({\it i.e.} 
relative humidity) and temperature.

It is useful to embark on a slight digression
from the analysis to consider several possible definitions of altitude.
Including density altitude, there are at least four possible types of
altitudes used in aerodynamic studies: geometric, geopotential, density,
and pressure.  Focus herein will be primarily on the first three.

Both geometric and geopotential altitudes are determined independently
of any atmospheric considerations.  Geometric altitude is literally the
physical elevation measured from mean sea level to a point above the surface.
Geopotential altitude could just as easily be defined as {\it equipotential}
altitude.  It is defined as the radius of a specific (gravitational) 
equipotential surface which surrounds the earth.  

Density and pressure altitude, on the other hand, are defined as the altitudes
which yields the standard altitude for a given set of parameters.  
The variation of pressure as a function of height $z$ is
\beq
\frac{dp}{dz} = -\rho(z) g(z) = -\frac{p g(z)}{RT(z)}
\eeq
which can be rearranged to give the integrals
\beq
\int_{p_0}^{p(H)} \frac{dp}{p} = -\frac{1}{R} \int_{0}^{H} 
\frac{dz\; g(z)}{T(z)}~.
\label{int0}
\eeq
For the small altitude changes considered herein, it is more than appropriate
to adopt the approximations $g(z) \approx g \equiv{\rm constant}$ and the 
linear temperature gradient $T(z) = T_0 - \Lambda z$, where 
$\Lambda$ is the temperature lapse rate.
In this case, Equation~\ref{int0} may be solved to give

\bea
\ln\left(\frac{p(H)}{p_0}\right)&=& \frac{g}{R\Lambda} 
  \ln\left(\frac{T_0}{T(H)}\right) \nonumber \\
 & = & \frac{g}{R\Lambda} \ln\left(\frac{T_0}{T_0 - \Lambda H}\right)~.
\label{int1}
\eea
After some mild algebra, the pressure altitude $H_p$ is determined 
to be

\beq
H_p \equiv \frac{T_0}{\Lambda} \left\{1-\left(
   \frac{p(H)}{p_0}\right)^\frac{R\Lambda}{g}\right\}
\label{p_alt}
\eeq
Since it is of greater interest to find an expression for altitude as a
function of air density, substituting $p(z) = \rho(z) R T(z)$ in 
Equation~\ref{int1} yields 

\beq
H_\rho = \frac{T_0}{\Lambda} \left\{ 1 - \left(\frac{RT_0 \rho(H)}{\mu P_0}\right)^{\left[\frac{\Lambda R}{g\mu-\Lambda R}\right]} \right\}
\label{dens_alt}
\eeq
which is the density altitude.

Here, $P_0$ and $T_0$ are the standard sea level pressure and
temperature, $R$ the gas constant, $\mu$ the molar mass of dry 
air, $\Lambda$ the temperature lapse rate, and $g$ the sea level
gravitational constant.  
The explicit values of these parameters are given in \cite{isa} as
\bea
T_0 & = & 288.15~{\rm K} \nonumber \\
P_0 & = & 101.325~{\rm kPa} \nonumber \\
g & = & 9.80665~{\rm ms}^{-2} \nonumber \\
\mu & = & 2.89644 \times 10^{-2}~{\rm kg}/{\rm mol} \nonumber \\
\Lambda & = & 6.5 \times 10^{-3}~{\rm K\cdot m}^{-1} \nonumber \\
R & = & 8.31432~{\rm J}\;{\rm mol}^{-1}\;{\rm K}^{-1} \nonumber
\eea

In the presence of humidity, the density $\rho$ is a combination
of both dry air density $\rho_a$ and water vapor density $\rho_v$. This
can be written in terms of the associated pressures as \cite{murray}
\beq
\rho = \rho_{a} + \rho_{v} = \frac{P_{a}}{R_{a}\;T} + \frac{P_v}{R_v \; T}~,
\eeq
where each term is derived from the ideal gas law, with $P_a$ and $P_v$ the
pressure of dry air and water vapor, and $R_a = 287.05$ and $R_v = 461.50$
the corresponding gas constants.  Since the total pressure is simply the sum
of both dry air pressure and vapor pressure, $P = P_a + P_v$, the
previous equation may be simplified to give
\beq
\rho = \frac{P-P_v}{R_a\; T} + \frac{P_v}{R_v \; T}~.
\eeq
                                                                                
The presence of humidity lowers the density of dry air, and hence lowers
aerodynamic drag.  However, the magnitude of this reduction is critically 
dependent on temperature.  A useful and generally accurate approximation 
to calculate vapor pressure is known as the Magnus Teten equation,
given by \cite{murray}
\beq
P_v \approx 10^{7.5 T/(237.7+T)}\cdot \frac{H_r}{100}
\eeq
where $H_r$ is the relative humidity measure and $T$ is the temperature
in $\dc$.

\begin{figure}[h]
\begin{center} \leavevmode
\includegraphics[angle=0,width=1.0\textwidth]{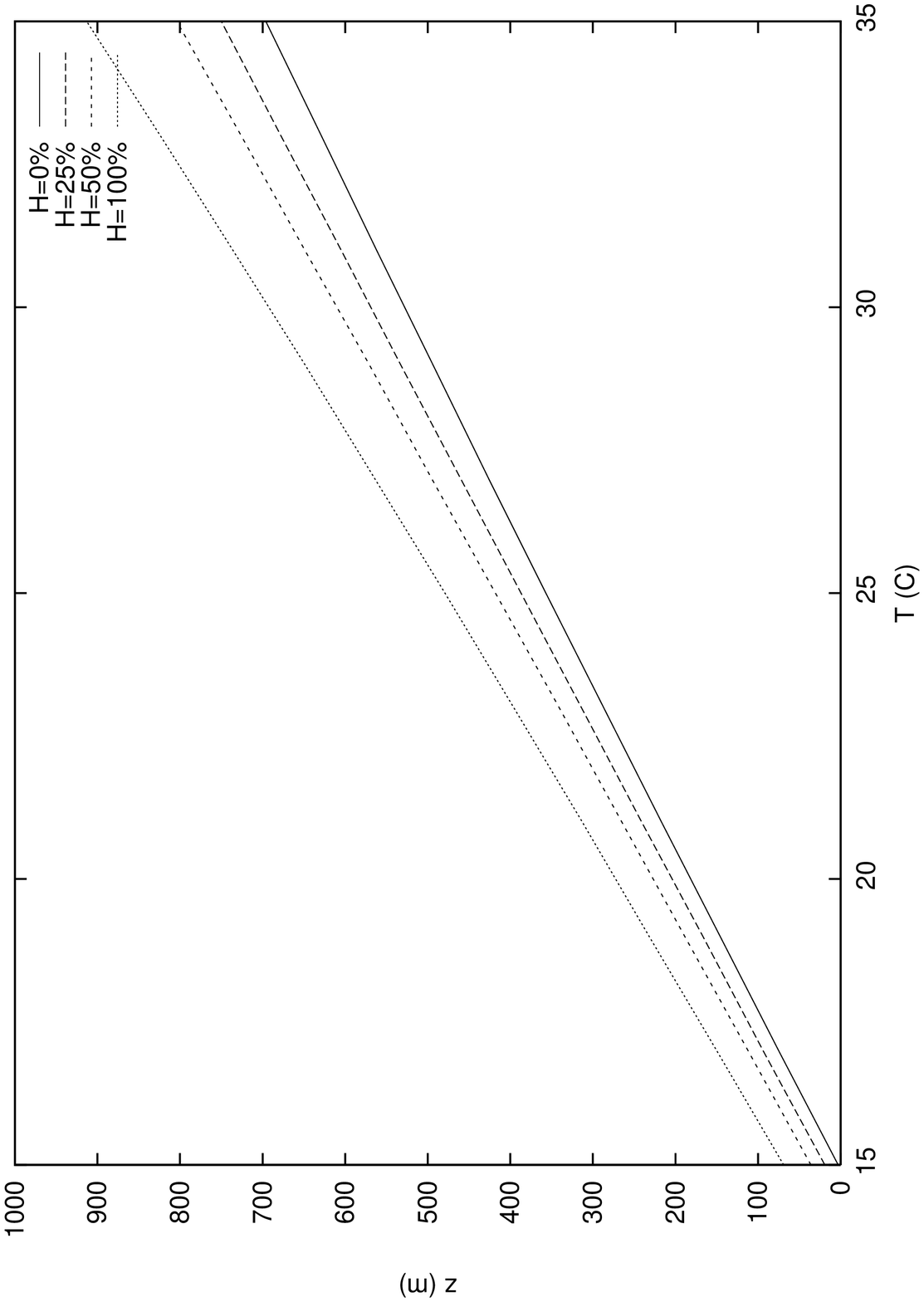}
\caption{Variation of density altitude as a function of temperature and
and relative humidity with barometer reading $P = 101.3~$kPa.}
\label{FigDA1}
\end{center}
\end{figure}
\begin{figure}[h]
\begin{center} \leavevmode
\includegraphics[angle=0,width=1.0\textwidth]{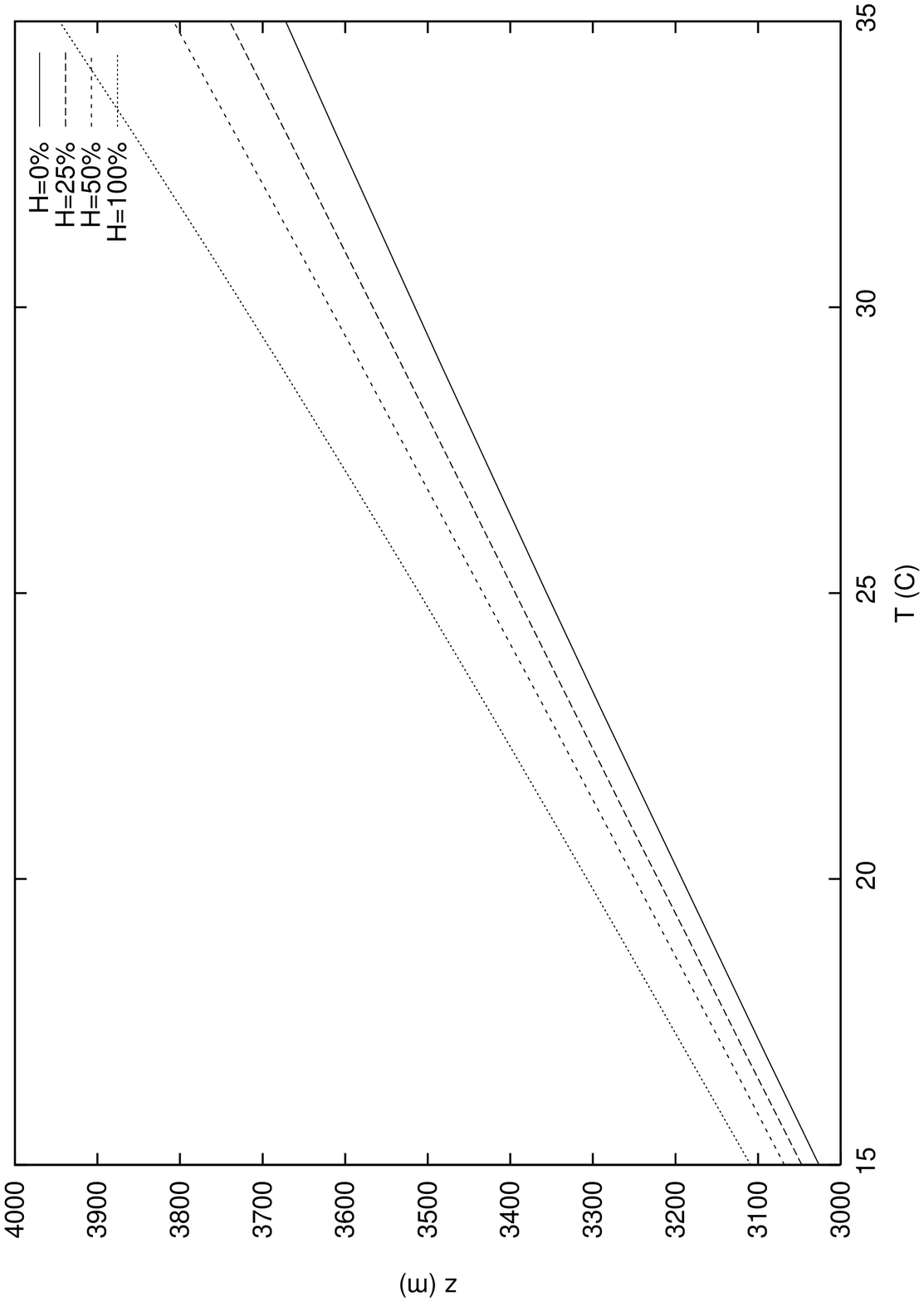}
\caption{Variation of density altitude as a function of temperature and
and relative humidity with barometer reading $P = 75~$kPa.}
\label{FigDA2}
\end{center}
\end{figure}

Figures~\ref{FigDA1} and
~\ref{FigDA2} show the calculated density altitudes for the temperature
range in question, including curves of constant relative humidity, for 
normal pressure (101.3~kPa) as well as high-altitude pressure (75~kPa). 
The latter is representative of venues such as those in Mexico City, where
world records were set in every sprint race and jumping event at the 1968
Olympic Games.

It is somewhat striking to note that a 20$\dc$ temperature range
at a fixed pressure can yield an effective altitude change of over 600~m, 
and even greater values with the addition of humidity and lowered station
pressure.  This suggests that
a combination of the effects considered in this paper can have significant
influence on sprint performances (especially in the 200~m which has been
shown to be strongly influenced by altitude \cite{jrm200}).

\subsection{A note on meteorological pressure}
A review of reported meteorological pressure variations seems to suggest 
that annual barometric averages for all cities range between about 
100-102~kPa.  This might
seem contradictory to common sense, since one naturally expects atmospheric
pressure to drop at higher altitudes.  Much in the same spirit as this work,
it is common practice to report {\it sea-level corrected} (SL)
pressures instead
of the measured (station) atmospheric pressures, in order to compare the
relative pressure differences experienced between weather stations, as
well as to predict the potential for weather system evolution.

Given a reported SL pressure $P_{\rm SL}$ at a (geophysical)
altitude $z$, the station pressure $P_{\rm stn}$ can be
computed as \cite{llnl}
\beq
P_{\rm stn} = P_{\rm SL} \left(\frac{288-0.0065z}{288}\right)^{5.2561} ~.
\eeq
The present study uses only station atmospheric pressures, but a conversion
formula which can utilize reported barometric pressure
will be offered in a subsequent section. 

\section{Results}
\label{results}
The model parameters used for the 100~meter analysis in this study are 
identical to those in \cite{jrm100},
\bea
f_0=6.10~{\rm ms}^{-2}&f_1 = 5.15~~{\rm ms}^{-2}&\sigma=2.2~{\rm s}^{-2} \nonumber \\
c=0.0385~{\rm s}^{-1}~&\alpha=0.323~{\rm s}^{-1}~~&A_d = 0.00288~{\rm m}^2{\rm kg}^{-1}
\label{params1}
\eea
as well as the ISA parameters described previously.  By definition, these
give a density altitude of 0~meters at 15$\dc$, $P_0 = 101.325~$kPa
and $0\%$ relative humidity.

Since few track meets are held at such ``low'' temperatures, the standard 
performance to which all others will be compared will be for the conditions
$P = 101.3~$kPa, $T=25\dc$, and a relative humidity level of $50\%$,  
which produces a raw time ({\it i.e.} excluding reaction) of 9.698~s.  Note 
that the corresponding density altitude for this performance
under the given conditions would be $H_\rho = 418$~meters. Thus, if the
performance were physically at sea level, the conditions would replicate
an atmosphere of elevation $H_\rho$.
Also, the relative humidity level is never measured at 0\%.  The
range of possible relative humidity readings considered herein are thus
constrained between 25\% and 100\%.  

\begin{figure}[h]
\begin{center} \leavevmode
\includegraphics[angle=0,width=1.0\textwidth]{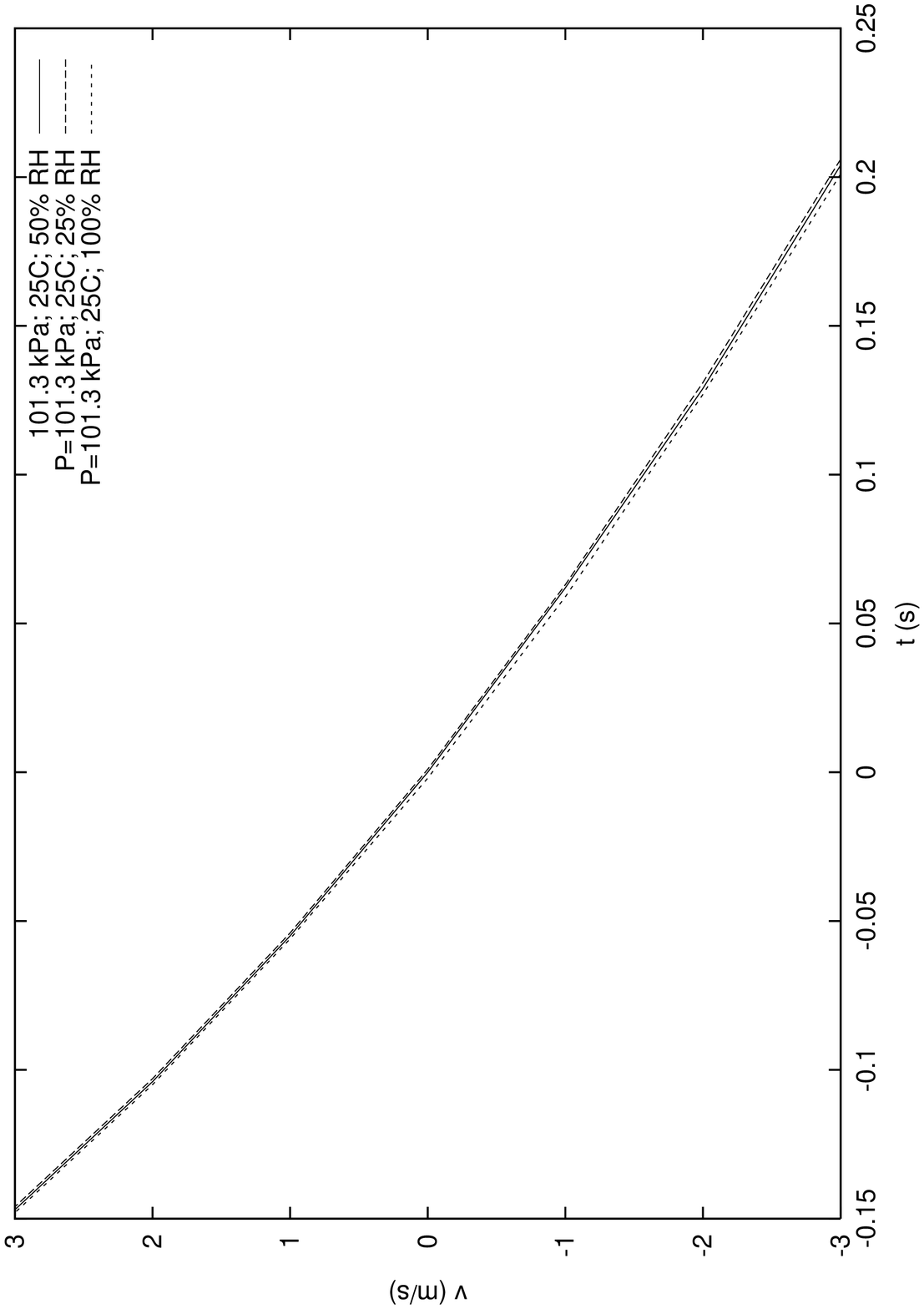}
\caption{Corrections to world class 100~meter sprint times for humidity 
variations between 25\% and 100\% RH at fixed pressure
(101.3~kPa) and temperature (25$\dc$).  The ``base'' time is 9.698~s 
in still conditions ($w=0\ms$) at sea level, $P = 101.3~$kPa,
$T = 25\dc$, and 50\% relative humidity (show as solid line).  
Negative corrections indicate
the performance was faster due to the ambient conditions, while 
positive corrections indicate a slower race time.}
\label{Fig100a}
\end{center}
\end{figure}

\begin{figure}[h]
\begin{center} \leavevmode
\includegraphics[angle=0,width=1.0\textwidth]{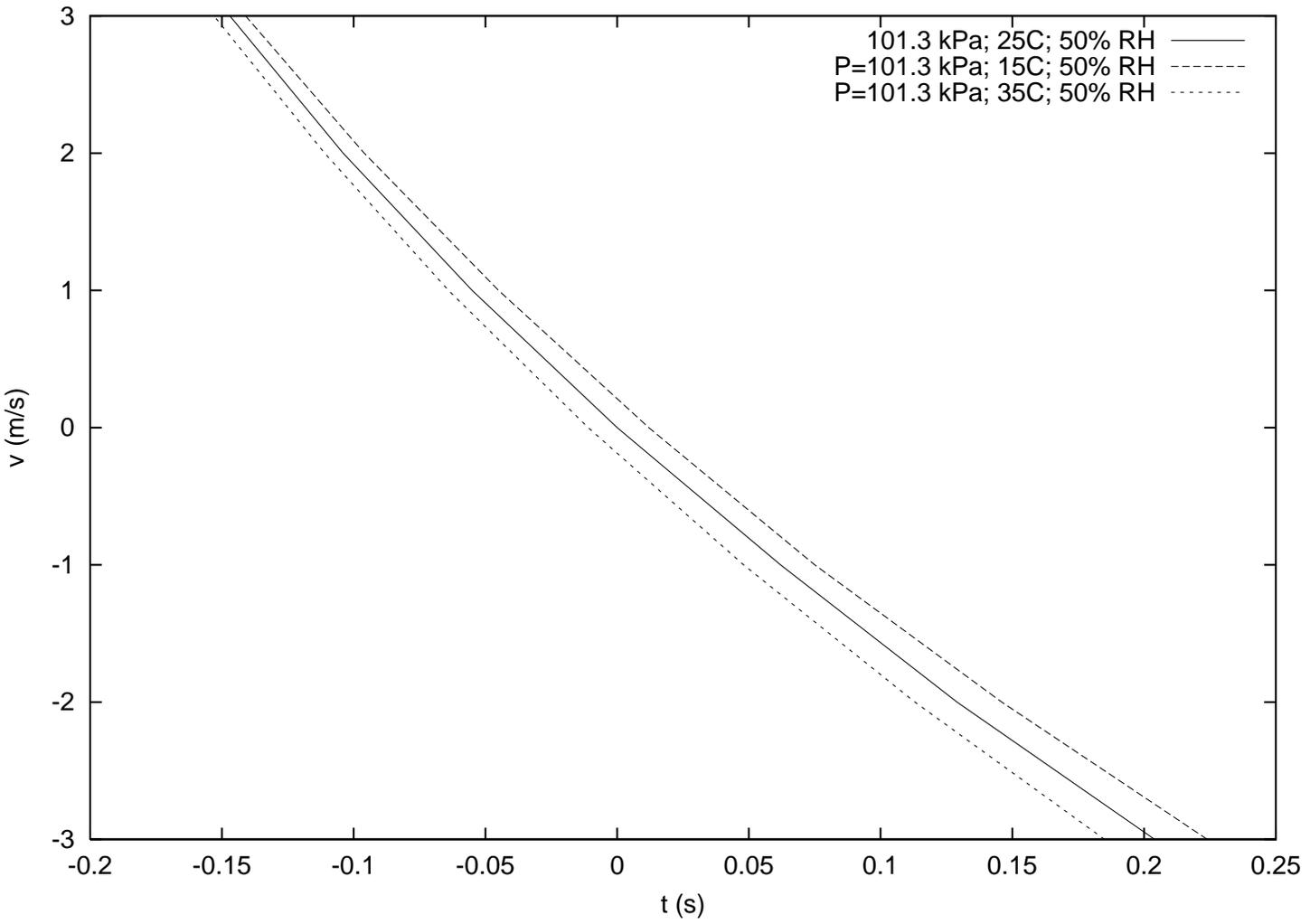}
\caption{Corrections to world class 100~meter sprint times for temperature variations between 15$\dc$ and 35$\dc$ at fixed barometric pressure (101.3~kPa) and relative humidity (50\%). }
\label{Fig100b}
\end{center}
\end{figure}

\begin{figure}[h]
\begin{center} \leavevmode
\includegraphics[angle=0,width=1.0\textwidth]{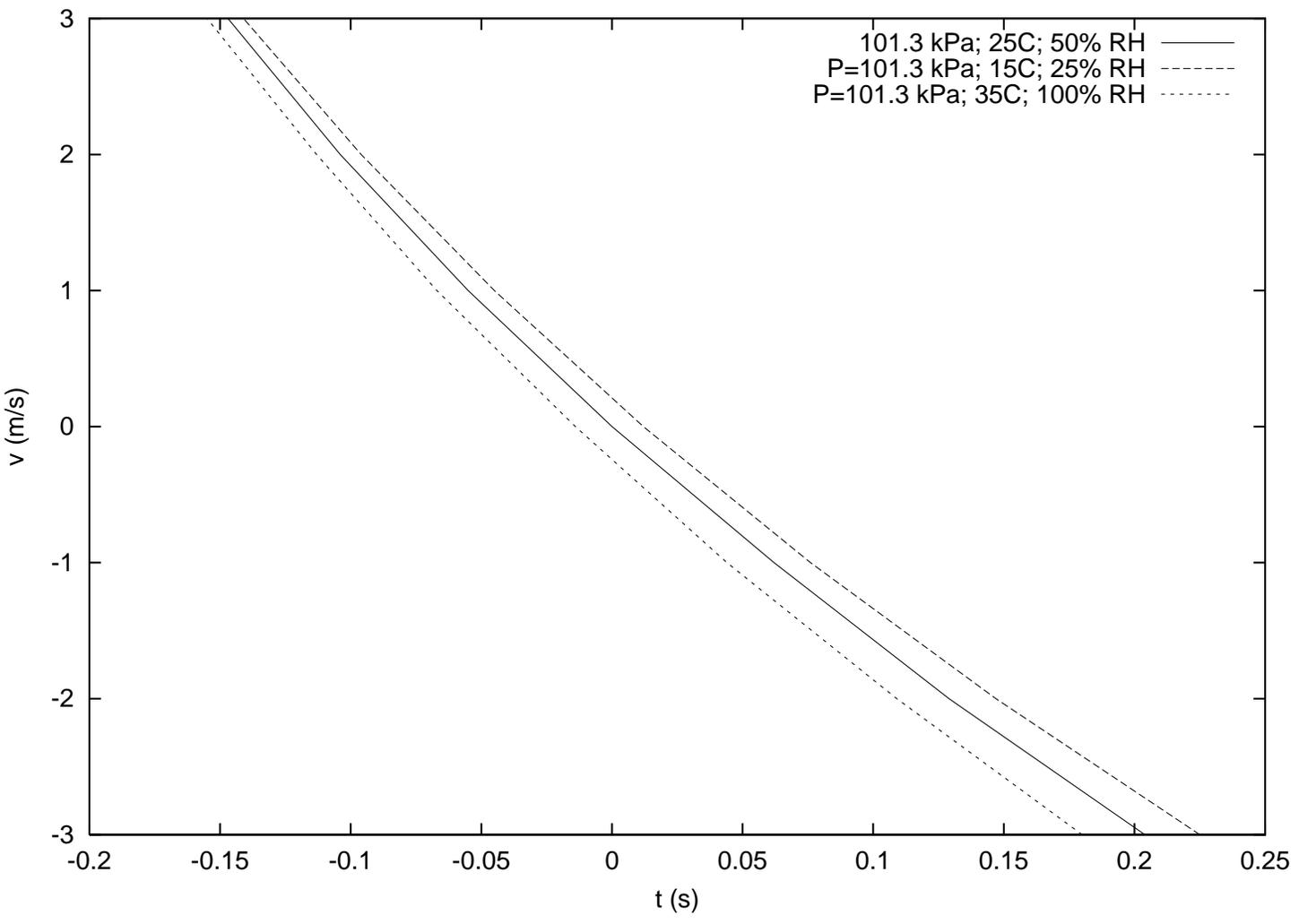}
\caption{Corrections to world class 100~meter sprint times for temperature and humidity variation at fixed barometric pressure (101.3~kPa)}
\label{Fig100c}
\end{center}
\end{figure}

\begin{figure}[h]
\begin{center} \leavevmode
\includegraphics[angle=0,width=1.0\textwidth]{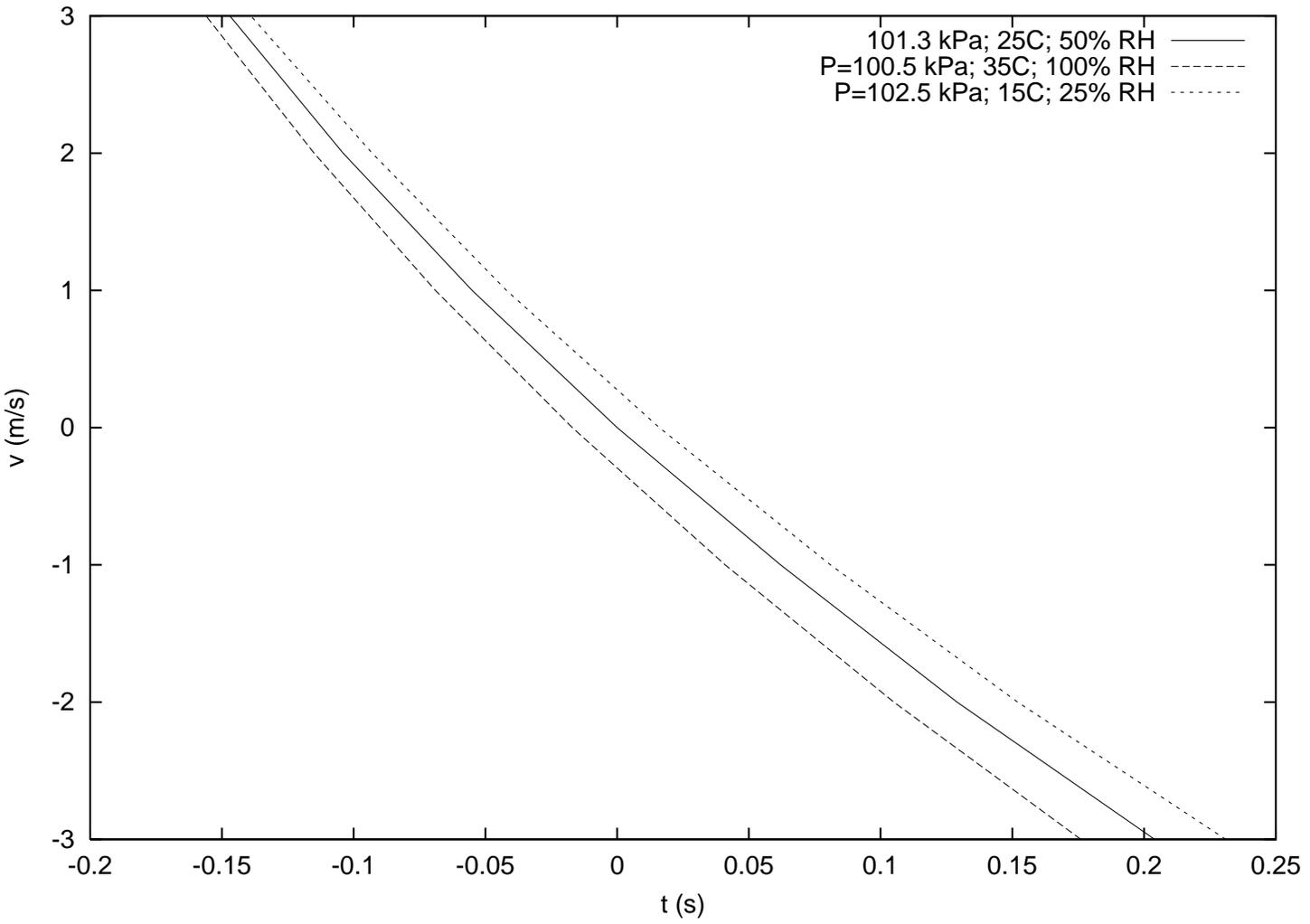}
\caption{Corrections to world class 100~meter sprint times for extremal 
pressure and temperature variations at the ``same'' venue (100.5~kPa, 35$\dc$, 100\% RH) and (102.5~kPa, 15$\dc$, 25\% RH).}
\label{Fig100d}
\end{center}
\end{figure}

\begin{figure}[h]
\begin{center} \leavevmode
\includegraphics[angle=0,width=1.0\textwidth]{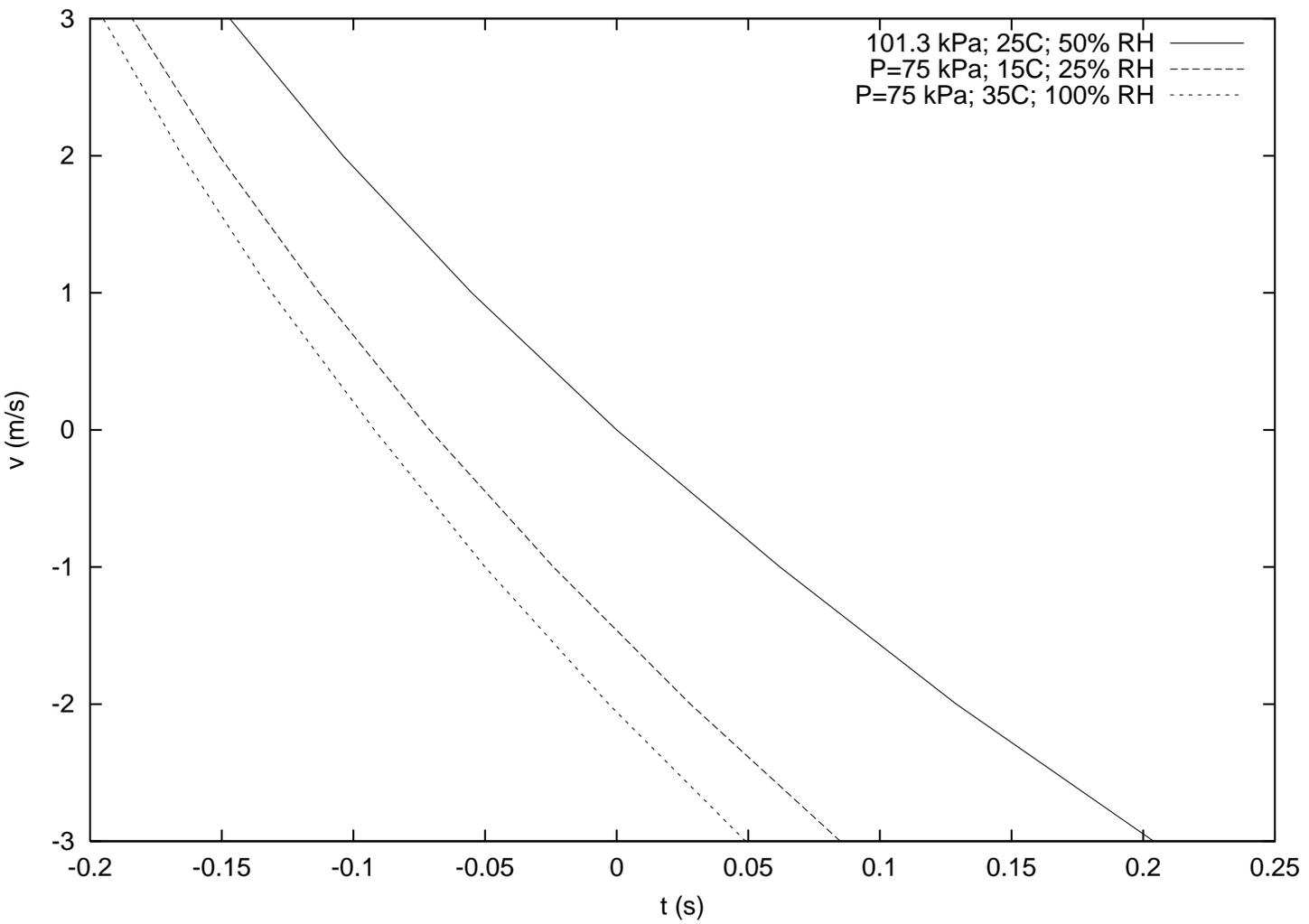}
\caption{Corrections to world class 100~meter sprint times for extremal 
condition variation at low pressure (75~kPa).}
\label{Fig100e}
\end{center}
\end{figure}

The associated performance adjustments in the 100~meter sprint for various 
conditions are displayed
in Figures~\ref{Fig100a} through \ref{Fig100f}, while Figures~\ref{Fig200a} 
through ~\ref{Fig200e} show possible adjustments to the 200~meter sprint
(see Section~\ref{section200} for further discussion).  General features of 
all graphs include a narrowing parameter space for increasing wind speeds.  
This implies
that regardless of the conditions, stronger tail winds will assist performances
by a smaller degree with respect to the ``base'' performance than will head
winds of similar strength.  This is completely consistent with previous 
results in the literature, and furthermore is to be expected based on the
mathematical form of the drag component.

Ranked in terms of magnitude of effect, humidity variations show the least
impact on race times.  Small changes in atmospheric pressure that one might
expect at a given venue show slightly more influence on times.  Temperature
changes over the range considered show the greatest individual impact.
However, it is the combination of the three factors that creates the greatest
impact, as is predictable based on the calculated density altitudes.

\subsection{Temperature and relative humidity variations}
At fixed pressure and temperature, the range of realistic humidity variations
shows little influence on 100~meter race times (Figure~\ref{Fig100a}), yielding
corrections of under 0.01~s for the range considered.  Since race times are
measured to this precision, the effects would be no doubt negligible.
The corrections for low pressure regions are also less than the 0.01~s.
Due to the extremely slim nature of the parameter space, the effects of
wind will essentially be the same regardless of the relative humidity reading.

Similarly, if temperature is allowed to vary at a fixed humidity level,
the corrections grow in magnitude but are still relatively small 
(Figure~\ref{Fig100b}).
Although one would realistically expect variations in barometer 
reading depending on the venue, the chosen values are to reflect a sampling
of the possible range of pressures recorded at the events.  On its own, 
temperature does not have a profound impact on the simulation times over 
the $\dT = 20\dc$ range considered herein.  With no wind or relative humidity
at fixed pressure (101.3~kPa), 
a 100~m performance will vary only 0.023~s in the given temperature range.
Overall, this corresponds to approximately a 750~m change in density 
altitude (the 15$\dc$ condition corresponds roughly to the standard 
atmosphere except for the non-zero humidity, and
thus a density altitude not significantly different from 0~m).  
Compared to the standard performance, 
temperatures below $25\dc$ are equivalent
to running at a lower altitude ({\it e.g.} below sea level).

A $+2\ms$ tailwind and standard conditions will
assist a world class 100~m performance by 0.104~s using the input parameters
defined in Equation~\ref{params1}.  This is essentially identical to the 
prediction in \cite{jrm100}, since the standard 100~m performance has been 
defined to be under these conditions.  Increasing the temperature to 35$\dc$ 
yields a time differential of $0.111$~seconds over the standard race, whereas 
decreasing the temperature to 15$\dc$ shows a difference of 0.097~seconds.
Hence, the difference in performance which should be observed over the
20$\dc$ range is roughly $0.02~$s.

In combination, these factors expand the parameter 
space from that of temperature variation alone.   Figure~\ref{Fig100c}
shows the region bounded by the lowest density altitude conditions 
(low temperature and low humidity) and the highest (higher temperature
and higher humidity).  In this case, with no wind the performances can
be up to 0.026~s different (effectively the additive result of the individual
temperature and humidity contributions).  Although such adjustments to the
performances seem small, it should be kept in mind that a difference of
0.03~s is a large margin in the 100~meter sprint, and can cause an athlete
to miss a qualifying time or even a record.

\subsection{Pressure variations}
\label{pvar100}
As one would expect, the greatest variations in performances next to wind
effects is introduced by changes in atmospheric pressure.  Barometric pressure
changes will be addressed in two ways.  First, variations in pressure for
a fixed venue will be considered.  Diurnal and seasonal fluctuations in
barometer reading are generally small for a given location \cite{ncdc}, 
at most 1~kPa from average.  

Model simulations for performances run at unusually high pressure (102.5~kPa) 
and unusually low pressure (100.5~kPa) with constant temperature and relative
humidity actually show little variation (under 0.01~s).  However, if the
temperature and humidity are such as to allow extreme under-dense and
over-dense atmospheres, then the effects are amplified.  Figure~\ref{Fig100d}
demonstrates how such ambient weather could potentially affect performances
run at the same venue.  With no wind, the performances can be
different by almost 0.03~seconds for the conditions considered, similar
to the predicted adjustments for the combined temperature and humidity
conditions given in the previous subsection.

The performance range for races run at high-altitude venues under similar
temperature and humidity extremes is plotted in Figure~\ref{Fig100e}.  The
base performance line is given as reference for the influence of larger
pressure changes.  At the lower pressure, the width of the
parameter space for zero wind conditions is 0.021~seconds, but the lower
density altitude point is already 0.071~s faster than the base performance.
Thus, depending on the atmospheric conditions at the altitude venue races
could be upward of 0.1~seconds faster than those run under typical conditions
at sea level.

\subsection{Combined effects of temperature, pressure, and humidity}
\begin{figure}[h]
\begin{center} \leavevmode
\includegraphics[angle=0,width=1.0\textwidth]{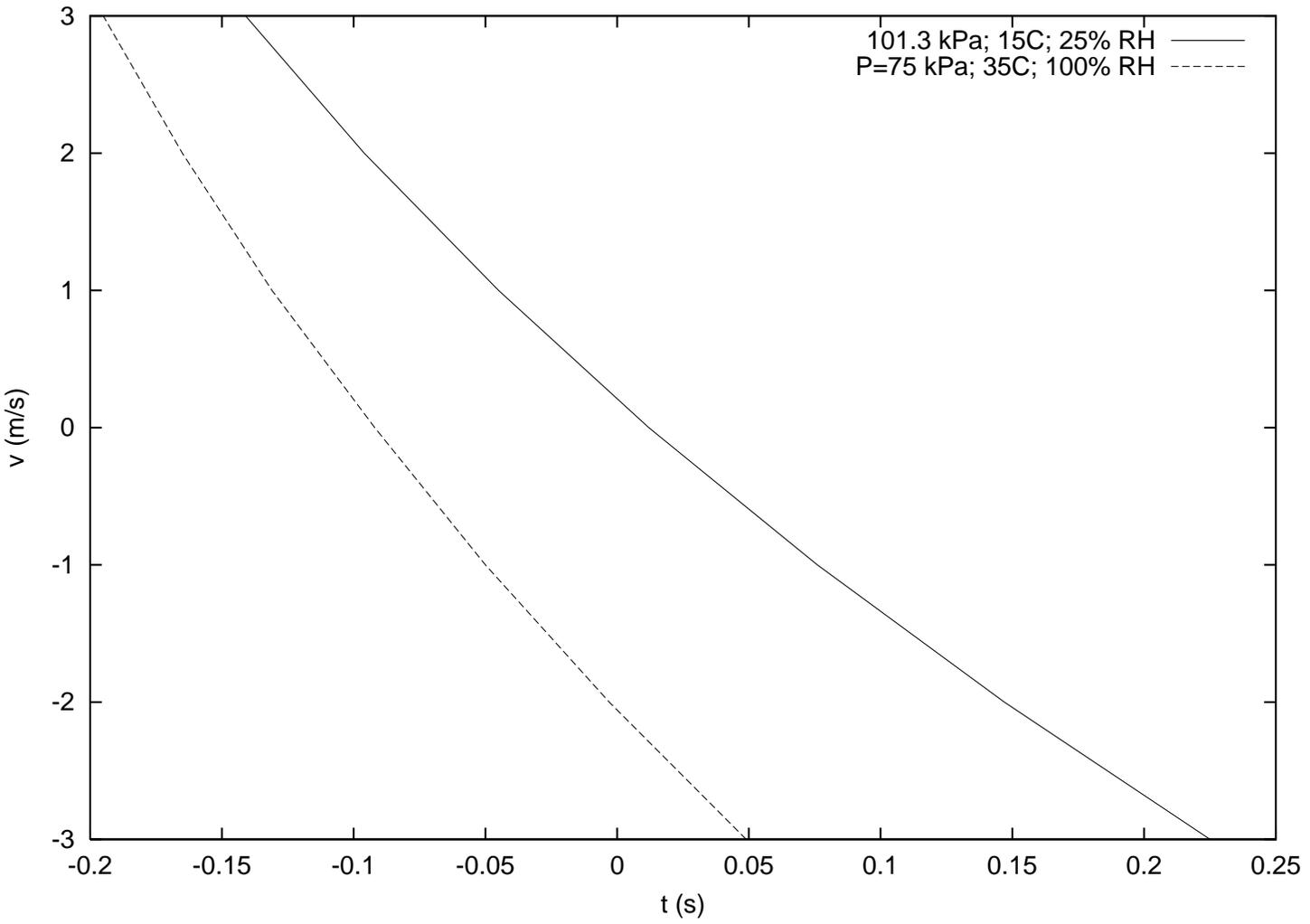}
\caption{Total range of 100~meter performance corrections for race times
held in all probable conditions and venues. }
\label{Fig100f}
\end{center}
\end{figure}

\begin{figure}[h]
\begin{center} \leavevmode
\includegraphics[angle=0,width=1.0\textwidth]{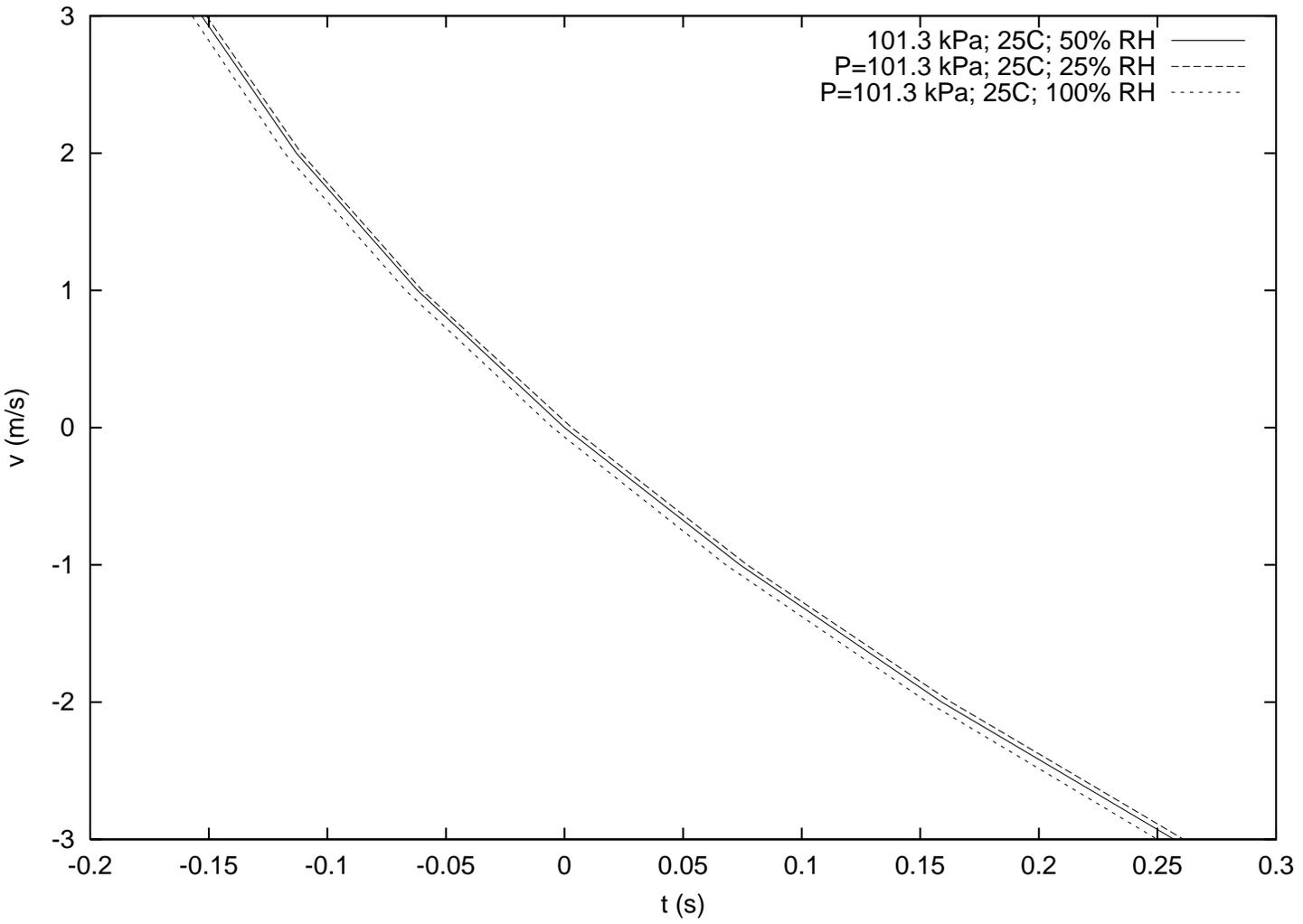}
\caption{Corrections to world class 200~meter sprint times for humidity 
variations at fixed temperature (25$\dc$) and pressure (101.3~kPa).  }
\label{Fig200a}
\end{center}
\end{figure}

\begin{figure}[h]
\begin{center} \leavevmode
\includegraphics[angle=0,width=1.0\textwidth]{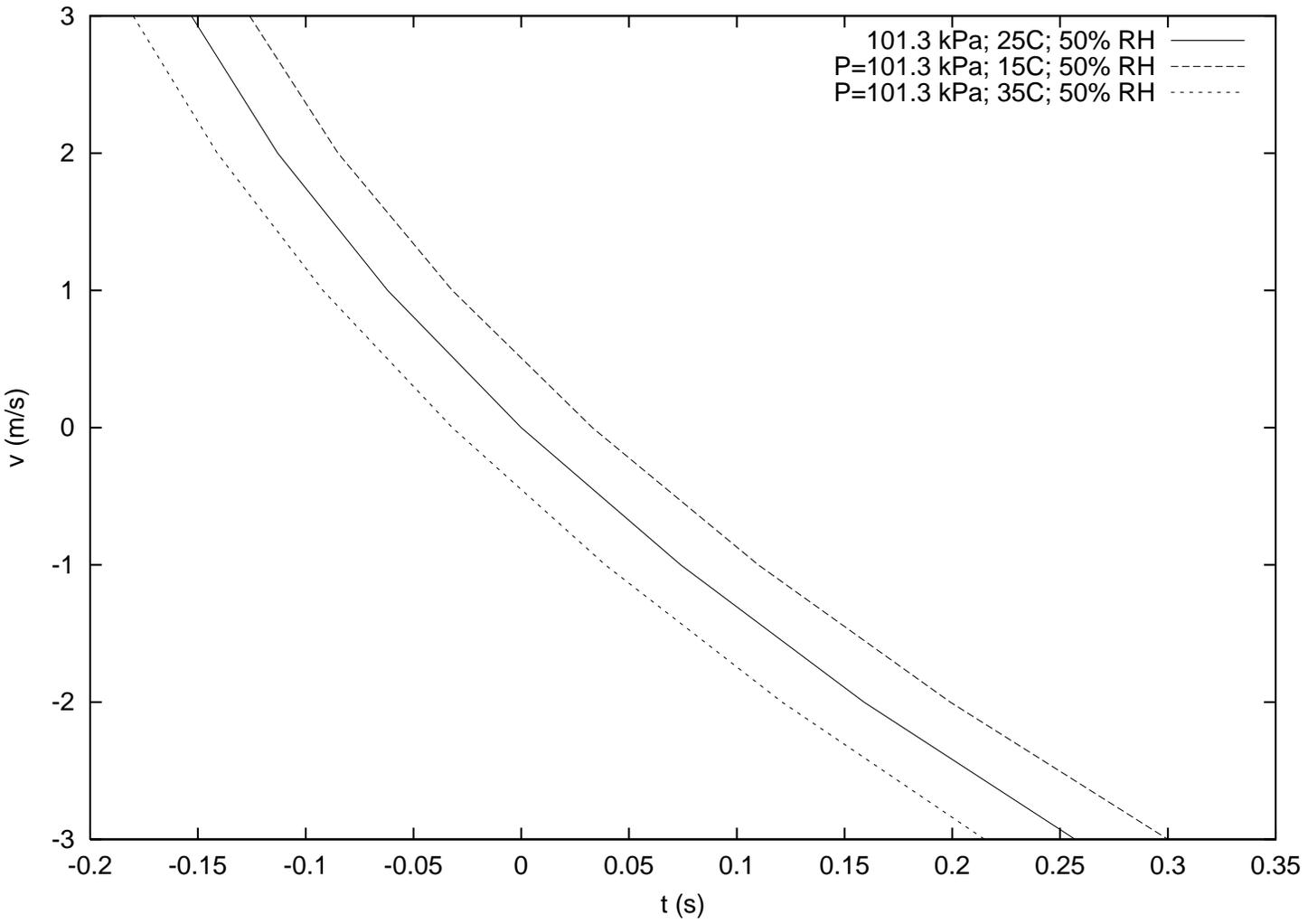}
\caption{Corrections to world class 200~meter sprint times for temperature 
variations at fixed pressure (101.3~kPa) and relative humidity (50\%).  }
\label{Fig200b}
\end{center}
\end{figure}

\begin{figure}[h]
\begin{center} \leavevmode
\includegraphics[angle=0,width=1.0\textwidth]{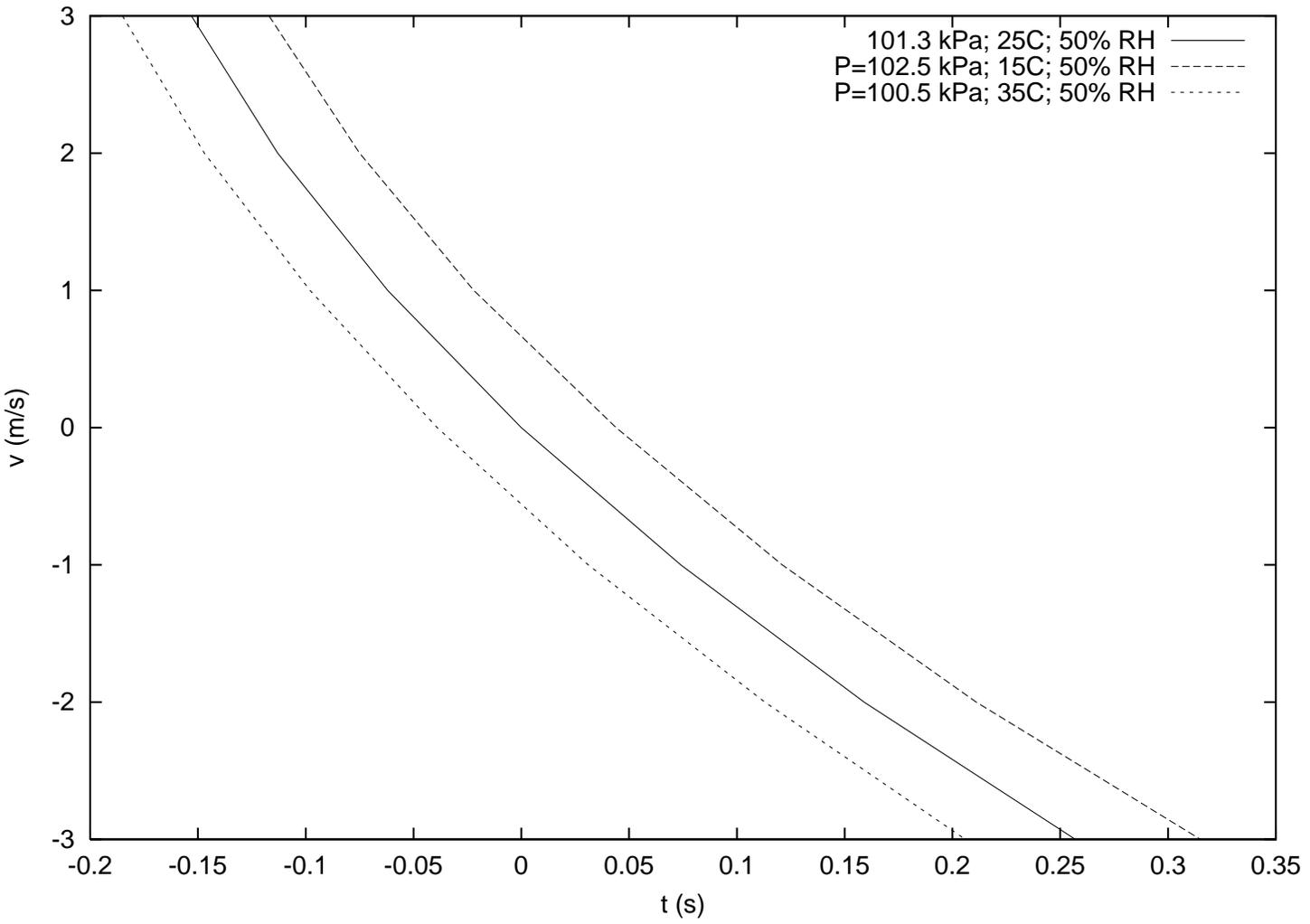}
\caption{
Corrections to world class 200~meter sprint times for barometric
pressure variation at a specific venue with fixed temperature (25$\dc$)
and relative humidity (50\%).
}
\label{Fig200c}
\end{center}
\end{figure}

\begin{figure}[h]
\begin{center} \leavevmode
\includegraphics[angle=0,width=1.0\textwidth]{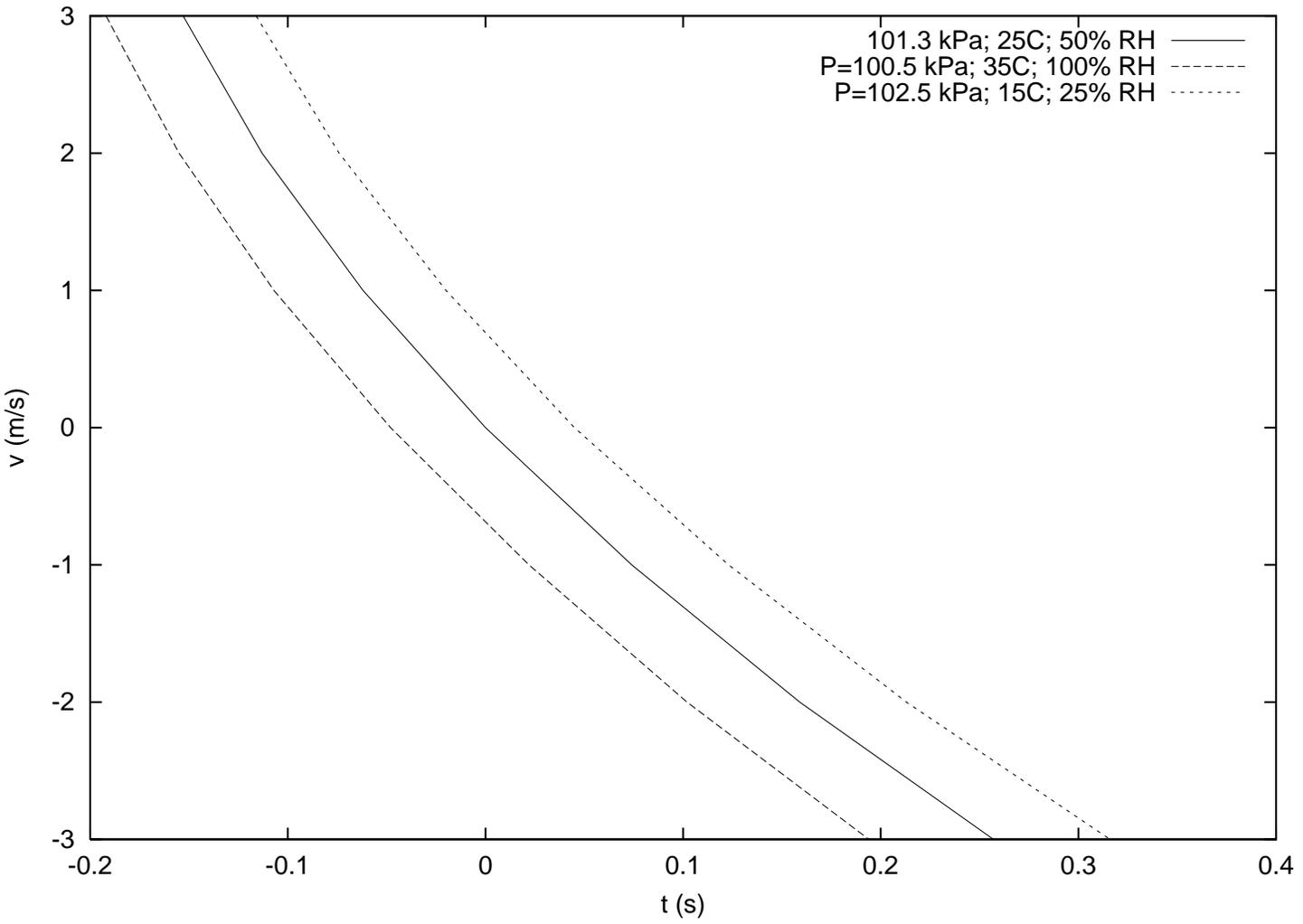}
\caption{
Corrections to world class 200~meter sprint times for 
extremal conditions at the ``same'' venue, (100.5~kPa, $35\dc$, 100\% RH)
and (102.5~kPa, $15\dc$, 25\% RH).
}
\label{Fig200d}
\end{center}
\end{figure}

The largest differences in performances arise when one considers combined
effects of pressure, temperature, and relative humidity.  Figure~\ref{Fig100f} 
shows the allowed parameter space regions for performances
run in extremal conditions: 75~kPa, 100\% humidity, and 35$\dc$ (yielding
the least dense atmosphere) and 101.3~kPa, 25\% humidity, 15$\dc$.  The
total allowed performance space is shown in Figure~\ref{Fig100f}.

Note that the difference in extremes is exceptionally amplified for head-winds.
Based on the total horizontal width in the Figure, a performance
run with a strong tail-wind (top of left curve) at high altitude can be
almost half a second faster than the same performance run with an equal-magnitude
headwind in extremely low density altitude conditions.

\section{Back-of-the-envelope conversion formula}
Reference~\cite{jrm100} gives a simple formula which can be used to correct
100~meter sprint times according to both wind and altitude conditions,
\begin{equation}
t_{0,0} \simeq  t_{w,H} [1.03 - 0.03 \exp(-0.000125\cdot H)
(1 - w\cdot t_{w,H} / 100)^2 ] ~.
\label{boteq}
\end{equation}
The time $t_{w,H}$ (s) is the recorded race time run with wind $w$ (ms$^{-1}$)
and at altitude $H$ (m), while the time $t_{0,0}$ is the adjusted time
as if it were run at sea level in calm conditions.
                                                                                
It is a simple task to modify Equation~\ref{boteq} to account instead for
density altitude.  The exponential term represents the change in altitude,
thus it can be replaced with a term of the form $\rho/\rho_0$, where
$\rho$ is the adjusted density, and $\rho_0$ is the reference density.
Alternatively, the original approximation in Equation~\ref{boteq} may be
used with the altitude $H$ being the density altitude.

Table~\ref{botecomp} shows how the top 100~meter performances are re-ordered
according to the given approximation.  These are compared with the original
back-of-the-envelope approximations given in Reference~\cite{jrm100}, based
exclusively on altitude.  The weather conditions have been taken from the
NC DC database \cite{ncdc}.  The relative humidity has been calculated from
the mean dew-point.  The cited temperature is the maximum temperature
recorded on that particular date, which is assumed to be reflective of
the conditions near the surface of the track (since the surface reflection
and re-emission of heat from the rubberized material usually increases the
temperature from the recorded mean).  Typical trackside temperatures reported 
in \cite{martin} support this argument.  For example, the 
surface temperature during the 100~m final in Atlanta (9.84~s, +0.7~$\ms$
performance in Table~\ref{botecomp}) was reported as 27.8$\dc$.

The effects of density altitude are for the most
part overshadowed by wind effects and are generally within 0.01~s of each
other after rounding to two decimal places. However, larger variations are 
observable in certain cases.  The most notable differences are in the
9.80~s performance (Maurice Greene, USA) at the 1999 World Championships in Seville, ESP, due
to the unusually high temperature, as well as the 9.85~s performance in
Ostrava, CZE (Asafa Powell, Jamaica).  The temperature measurement during the latter was reportedly
a chilly 10$\dc$ and humidity levels near 100\%.  

At the time of writing of this manuscript, the world record in the men's
100~meter sprint is 9.77~seconds by Asafa Powell, set on 14 June
2005 in Athens, Greece (note that although there is a faster performance
listed in the Table, only races with winds under +2$\ms$ are eligible for
record status).  The wind reading for this performance was $+1.6\ms$.
According to Table~\ref{botecomp}, Powell's world record run actually 
adjusts to about 9.85~s, very close to his earlier time from Ostrava.

In fact, prior to this the world record was
9.78~s by Tim Montgomery of the USA.  The wind in this race was just at
the legal limit for performance ratification ($+2\ms$).  This time adjusts
to an even slower performance of 9.87~s with density altitude considerations
(9.88~s using the older method).  Both of these times are eclipsed by the
former world record of Maurice Greene, who posted a time of 9.79~s
(+0.1$\ms$) in Athens roughly six years to the day prior to Powell's race.
This time adjusts to between 9.80-9.81~s depending on the conversion method.

\section{Corrections to 200~meter race times}
\label{section200}
Aerodynamic drag effects in the 200~m sprint have been found to be compounded
due to the longer duration and distance of the race.  
In Reference~\cite{jrm200}, it was suggested that ``extreme'' conditions such
as high wind and altitude can considerably affect performances for better or
worse.  That is, a 1000~m altitude alone can improve a world class sea-level 
performance by up to 0.1~s, the equivalent assistance provided by a $+2\ms$ 
tail-wind in the 100~m sprint.  Higher altitudes can yield even greater boosts. 
Thus, the variability of density altitude would seem to be all the more 
relevant to the 200~m sprint.

This section does not seek to provide a comprehensive analysis for the 200~m
({\it e.g.} the influence of cross-winds or lane dependence), so only the
results for a race run in lane~4 of a standard IAAF outdoor track will
be reported.  The model equations (\ref{quasi}) are adapted for the curve
according to Reference~\cite{jrm200},
\beq
\dot{v}(t) = \beta (f_s + f_m) - f_v - f_d~,
\eeq
where
\beq
\beta(v(t); R_l) = (1\; -\;\xi\;v(t)^2/R_{\cal l} )~, \label{damping}
\eeq 
and $R_l$ is the radius of curvature (in meters) for the track in lane ${\cal l}$,
\beq R_l = 36.80+1.22\;({\cal l} - 1)~, \eeq 
compliant with the standard IAAF track.
For the 200~m, slightly different parameters used are to reflect
the more ``energy-conservative'' strategy for the race \cite{jrm200}:
$f_0 = 6.0\;{\rm ms}^{-2}; f_1 = 4.83\;{\rm ms}^{-2}; c = 0.024~{\rm s}^{-1}$.
The other parameters remain unchanged.  The curve factor is $\xi = 0.015$.
As with the 100~meter sprint, the standard performance adopted for comparison
is 19.727~s, run in lane 4 at 25$\dc$ and 50\% relative humidity.  Under these
conditions, a 2$\ms$ wind assists the sprinter by approximately 0.113~s.

The effects of humidity alone are again effectively inconsequential, in this
case showing less than a 0.01~s differential with no wind (Figure~\ref{Fig200a}.
For wind speeds
of 2$\ms$ at the same pressure value combined with a  50\% increase in 
humidity, the advantage over the base conditions grows to 0.118~s.  Thus,
for all wind speeds considered, the effects of humidity are negligible.

Temperature plays a much stronger role in the longer sprint.
Figure~\ref{Fig200b} shows this effect for fixed pressure and relative humidity.
In both 15$\dc$ and 35$\dc$ weather, a world class 200~meter sprint can be
approximately 0.03~s slower or faster, giving a total differential of
0.065~s over the entire 20$\dc$ range.

Figure~\ref{Fig200c} demonstrates the effects of pressure variation at
a specific venue (at constant temperature and relative humidity).  At higher
pressures (102.5~kPa), the simulation shows the race time slowed by 
0.011~s, while at unusually low pressures (100.5~kPa) the race by only 0.007~s.

In extremal conditions, including temperature and humidity variations as well 
causes these differentials to dramatically change (Figure~\ref{Fig200d}.  
For unusually high
pressure, low temperature and humidity conditions, the race time is 
0.045~s slower as compared to the base conditions.  On the other hand,
the lower pressure, higher temperature and humidity conditions yield
a 0.048~s decrease in the race time.   Thus, the difference between two
200~m races run at the same venue but under these vastly differing 
conditions could up up to 0.1~seconds different even if there is no
wind present.

\begin{figure}[h]
\begin{center} \leavevmode
\includegraphics[angle=0,width=1.0\textwidth]{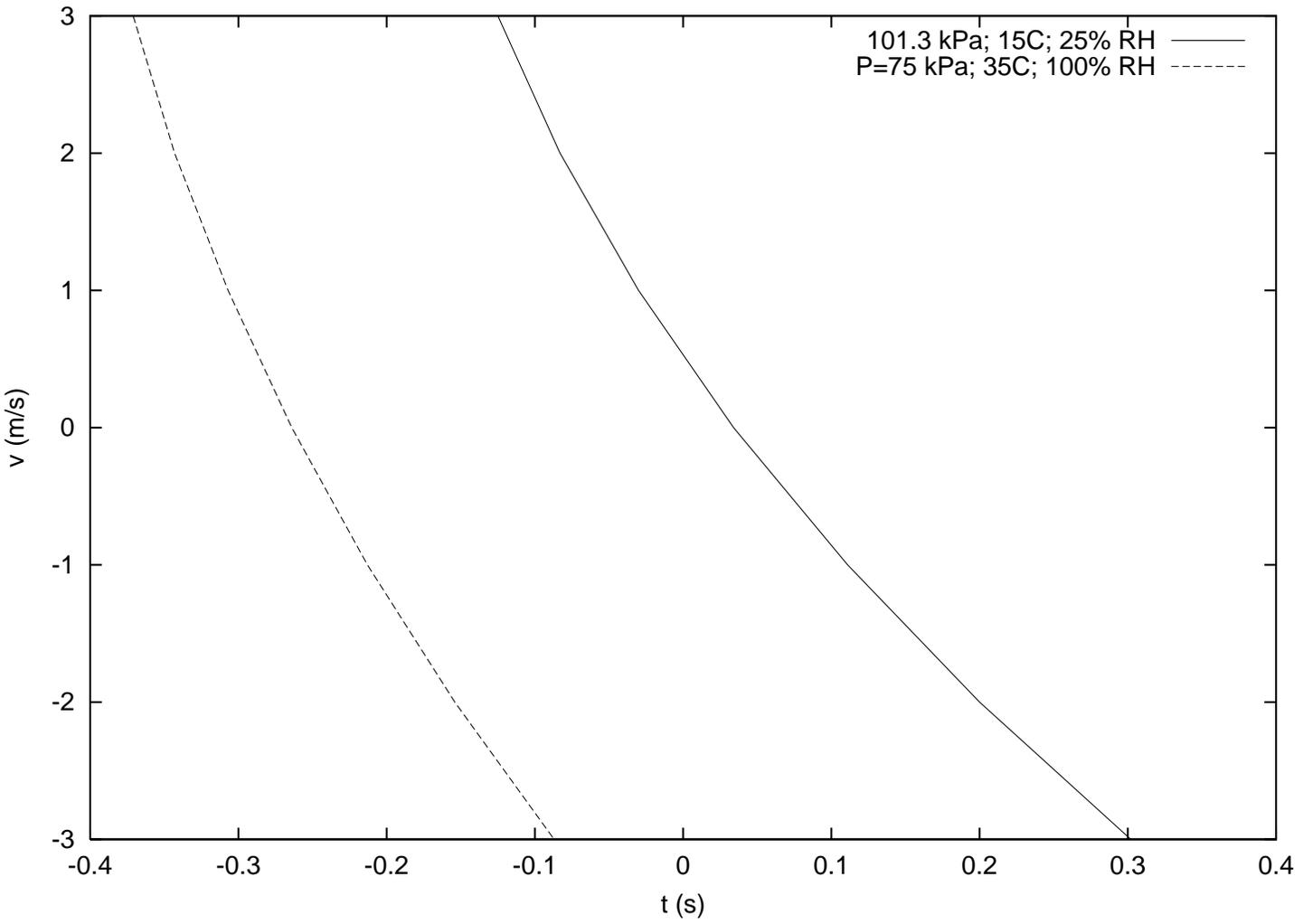}
\caption{ 
Corrections to world class 200~meter sprint times for 
extremal conditions at different venues, (75~kPa, $35\dc$, 100\% RH)
and (101.3~kPa, $15\dc$, 25\% RH).
}
\label{Fig200e}
\end{center}
\end{figure}

The most striking difference is observed when considering differences between
venues or large station pressure differences (Figure~\ref{Fig200e}).  
With no wind, the difference
between the base conditions and those at high altitude (at the same
temperature and relative humidity) can be
0.23~s, which is consistent with the figures reported in Reference~\cite{jrm200}
for a 2500~m altitude difference.  Recall that the difference in density 
altitude between these two conditions is roughly 3000~m.  

A 2$\ms$ wind improves the low pressure performance by only 0.083~s over still
conditions at the same pressure, but when compared to the base conditions
this figure becomes 0.313~s.  In fact, the complete horizontal span of the 
parameter space depicted in Figure~\ref{Fig200e} is almost 0.64~seconds,
demonstrating the exceeding variability that one could expect in 200~m
race times.  Again, this assumes that the wind is blowing exclusively in
one direction (down the 100~meter straight portion of the track), and is
unchanged throughout the duration of the race.  

\subsection{Can density altitude explain the men's 200~meter world record?}
\label{mjwr}
Michael Johnson (USA) upset the standard in the men's 200~m event in 1996
when he set two world records over the course of the summer.  At the USATF
Championships in Atlanta, his time of 19.66~s (wind $+1.6~\ms$) erased the
25~year old record of 19.72~s (set in Mexico City, and thus altitude assisted).
However, it was his performance of 19.32~s (wind $+0.4~\ms$) at the Olympic 
Games which truly redefined the race.  Reference~\cite{jrm200} offers a
thorough analysis of the race which addresses wind and altitude effects.
It was suggested that overall, the race received less than a 0.1~s boost from
the combined conditions (Atlanta sits at roughly 350~m above sea level).

In light of the current analysis, however, the question can be posed: ``By
how much {\it could} Johnson's race have been assisted?''.  Extensive
trackside meteorological data was recorded at the Games, and is reported
in Reference\cite{martin}.  
According to this data, the mean surface temperature during the race (21:00~EDT,
01 August 1996)
was 34.7$\dc$, with a relative humidity of 67\%.  From the NCDC database,
the adjusted sea-level pressure was recorded as 101.7~kPa.  Combined,
these values give a density altitude of 1175~meters (Atlanta's physical 
elevation is roughly 315~meters).

As compared to the base conditions (a density altitude of about 400~m), this 
represents an 800~m change in effective altitude.  According to the
results of Reference~\cite{jrm200}, such an altitude increase results in
an advantage of approximately 0.05~seconds.  When the minimal wind is taken
into account (0.3$\ms$), the difference rises to 0.1~s.  Previously,
the ``corrected'' value of the World Record was reported as 19.38~s 
\cite{jrm200} using only wind-speed and physical altitude, so the 
inclusion of density altitude enhances the correction by 0.04~s and
thus adjusts the time to a base 19.42~s.

While this does not explain the enormous improvement over the previous
record, is does highlight that density altitude considerations become 
increasingly important in the 200~meter sprint, and no doubt even moreso
for the 400~meter event.

\section{Conclusions}
This report has considered the effect of pressure, temperature, and relative
humidity variations on 100~ and 200~meter sprint performances.  It has been
determined that the influence of each condition can be ranked (in order of
increasing assistance) by humidity, pressure, and temperature.  The combined
effects of each are essentially additive.  When wind conditions are 
taken into account, the impact is amplified for head-winds but dampened for
tail-winds.

All in all, the use of density altitude over geophysical (or geopotential)
altitude seems somewhat irrelevant, since the associated corrections are
different by a few hundredths of a second at best.  Nevertheless, it is
the difference of these few hundredths which can secure a performance in
the record books, or lead to lucrative endorsement deals for world class
athletes.

\vskip 1cm

\noindent{\bf {\Large Acknowledgments}} \\
I thank J.\ Dapena (Indiana University) for insightful discussions.
This work has been financially supported by a grant from Loyola Marymount
University.

\pagebreak
\begin{table}[h]
\begin{tabular}{ccccccc}\hline
$t_w$ ($w$) & Venue ($H$) & Date & $T$, $P_{\rm SL}$, RH & $t_{\rm DA}$  & $t_{\rm PA}$ \\
 
9.69 (+5.7)& El Paso, TX (1300)& 13/04/96& 100.5, 25.6, 13 & 9.910 & 9.915 \\
9.77 (+1.6)& Athens, GRE (100)& 14/06/05& 101.5, 28.9, 36 & 9.853 & 9.851 \\
9.78 (+2.0)& Paris, FRA (50) &14/09/02& 102.2, 21.7, 39 & 9.873 & 9.877\\
9.79 (+0.1)& Athens, GRE (100)& 16/06/99& 101.7, 32.8, 38&9.804 & 9.799 \\
9.80 (+0.2)& Seville, ESP (0) &22/08/99& 101.3, 41.7, 23&9.824 & 9.811\\
9.82 (-0.2)& Edmonton, AB (700) &05/08/01& 101.2, 25, 40 & 9.831 & 9.832 \\ 
9.84 (+0.7)& Atlanta, GA (310) & 27/07/96 & 102.2, 27.2, 70 & 9.885 & 9.885 \\ 
9.85 (+0.6)& Ostrava, CZE (250) & 09/06/05 & 101.6, 10.0, 100 & 9.874 & 9.889 \\
10.03 (-2.1) & Abbotsford, BC (50) & 19/07/97 & 101.9, 27.1, 44 & 9.901 & 9.901 \\ \hline
\end{tabular}
\caption{Adjusted top performances using back-of-the-envelope conversion
algorithm for density altitude ($t_{\rm DA}$), as compared to 
physical altitude correction ($t_{\rm PA}$).  All times are expressed
in seconds (s), elevation $H$ in meters (m), temperature $T$ in $\dc$,
SL-pressure $P_{\rm SL}$ in kPa, and relative humidity (RH) in \%.}
\label{botecomp}
\end{table}

\pagebreak

\end{document}